\documentclass[11pt,letterpaper]{article}
\usepackage{graphicx}
\usepackage{bm}
\usepackage{amssymb}
\usepackage{color}
\usepackage{amsmath}
\usepackage[numbers]{natbib}
\usepackage{enumerate}
\usepackage{amstext}
\usepackage{footmisc}
\usepackage{latexsym}
\usepackage[usenames,dvipsnames]{xcolor}
\usepackage[colorlinks=true,citecolor=Cerulean,linkcolor=RubineRed,urlcolor=Cerulean]{hyperref}
\usepackage{multibib}
\usepackage{ fixfoot}
\usepackage{fancyhdr}
\newcommand{\pap}[1]{\left( #1 \right)}
\newcommand{\pas}[1]{\left[#1 \right]}

\DeclareFixedFootnote{\rep}{Electronic address: \href{mailto:fj.gomez34@dipc.org}{fj.gomez34@dipc.org}}

\newcommand{\beq}{\begin{equation}}
\newcommand{\eeq}{\end{equation}}
\newcommand{\beqa}{\begin{eqnarray}}
\newcommand{\eeqa}{\end{eqnarray}}

\begin{document}
\title{{\bf Experimentally testing quantum critical dynamics beyond the Kibble-Zurek mechanism}}
\author{Jin-Ming Cui$^{1,2}$, Fernando Javier G{\'o}mez-Ruiz$^{3,4,6}$, Yun-Feng Huang$^{1,2,*}$,\\
Chuan-Feng Li$^{1,2,\dagger}$, Guang-Can Guo$^{1,2}$ \& Adolfo del Campo$^{3,5,6,7,\ddagger}$}
\date{}
\maketitle
\vspace{-1cm}
\begin{center}
$^{1}${\it CAS Key Laboratory of Quantum Information, University of Science and Technology of China, Hefei 230026, China}\\
$^{2}${\it CAS Center For Excellence in Quantum Information and Quantum Physics, University of Science and Technology of China, Hefei 230026, China}\\
$^{3}${\it Donostia International Physics Center, E-20018 San Sebasti{\'a}n, Spain}\\
$^{4}${\it Departamento de F{\'i}sica, Universidad de los Andes, A.A. 4976, Bogot{\'a} D. C., Colombia}\\
$^{5}${\it  IKERBASQUE, Basque Foundation for Science, E-48013 Bilbao, Spain}\\
$^{6}${\it  Department of Physics, University of Massachusetts Boston, 100 Morrissey Boulevard, Boston, MA 02125}\\
$^{7}${\it Theoretical Division, Los Alamos National Laboratory, MS-B213, Los Alamos, NM 87545, USA}
\end{center}
\begin{abstract}
We experimentally probe the  distribution of kink pairs resulting from driving a one-dimensional quantum Ising chain through the paramagnet-ferromagnet quantum phase transition, using a single trapped ion as a quantum simulator in momentum space. The number of kink pairs after the transition follows a Poisson binomial distribution, in which all cumulants scale with a universal power-law as a function of the quench time in which the transition is crossed. We experimentally verified this scaling for the first cumulants and report deviations due to noise-induced  dephasing of the trapped ion. Our results establish that the universal character of the critical dynamics can be extended beyond the paradigmatic Kibble-Zurek mechanism, which accounts for the mean kink number, to characterize the full probability distribution of topological defects.
\end{abstract}

\section*{Introduction}
The understanding of nonequilibrium quantum matter stands out as a fascinating open problem at the frontiers of physics. Few theoretical tools account for the behavior away from equilibrium in terms of equilibrium properties. One such paradigm is the so-called Kibble-Zurek mechanism (KZM) that describes the nonadiabatic dynamics across a phase transition and predicts the formation of topological defects, such as vortices in superfluids and domain walls in spin systems~\cite{DZ14}. Pioneering insights on the KZM were conceived in a cosmological setting~\cite{Kibble76a} and applied to describe thermal continuous phase transitions~\cite{Kibble76b,Zurek96a,Zurek96c}. The resulting KZM was latter extended to quantum phase transitions~\cite{Polkovnikov05,Damski05,Dziarmaga05,ZDZ05}. Its central prediction is that the average total number of topological defects $\langle n^T\rangle$, formed when a system is driven through a critical point in a time scale $\tau_{Q}$, is given by a universal power-law $\langle n^T\rangle\sim\tau_{Q}^{-\beta}$. The power-law exponent $\beta=D\nu/(1+z\nu)$ is fixed by the dimensionality of the system $D$ and a combination of the equilibrium correlation-length and dynamic critical exponents, denoted by $\nu$ and $z$, respectively. Essentially, the KZM is a statement about the breakdown of the adiabatic dynamics across a critical point. As such, it provides useful heuristics for the preparation of ground-state phases of matter in the laboratory, e.g., in quantum simulation and adiabatic quantum computation~\cite{Suzuki09b}. It has spurred a wide variety of experimental efforts in superfluid Helium~\cite{Hendry1994,Ruutu1996,Bauerle1996}, liquid crystals~\cite{Chuang91,Bowick94}, convective fluids~\cite{Casado01,Casado06}, superconducting rings~\cite{Monaco02,Monaco06,Weir13},  trapped-ions~\cite{Ulm13,Pyka13,EH13}, colloids~\cite{Keim15}, and ultracold atoms~\cite{Weiler08,Lamporesi13,Chomaz15,Navon15,Ko19}, to name some relevant examples. This activity has advanced our understanding of critical dynamics, e.g., by extending the KZM to inhomogeneous systems~\cite{DKZ13,Fernando19}.
In the quantum domain, experimental progress has been more limited and led by the use of quantum simulators in a variety of platforms~\cite{Xu2014,Cui16,Wu16}.

Beyond the focus of the KZM, the full counting statistics encoded in the probability distribution  can be expected to shed further light. The number distribution of topological defects  has become available in recent experiments~\cite{Lukin17}. In addition, the distribution of kinks  has recently been explored theoretically in the one-dimensional transverse-field quantum Ising model (TFQIM)~\cite{delcampo18}, a paradigmatic testbed of quantum phase transitions~\cite{Sachdev}. Indeed, it has been  argued that  the distribution of topological defects is generally  determined by the scaling theory of phase transitions and thus exhibits signatures of universality beyond the KZM~\cite{delcampo18}. We aim at validating this prediction experimentally. 

The TFQIM is described by the Hamiltonian 
\begin{eqnarray}
\hat{\mathcal{H}}=-J\sum_{m=1}^{N} \pap{\hat{\sigma}_{m}^{z}\hat{\sigma}_{m+1}^{z}+g\hat{\sigma}_{m}^{x}},\label{H_Ising}
\end{eqnarray}
that we shall consider with periodic boundary conditions $\hat{\sigma}_{1}^{z}=\hat{\sigma}_{N+1}^{z}$ and $N$ even. Additionally, $\hat{\sigma}_{m}^{z}$ and $\hat{\sigma}_{m}^{x}$ are Pauli matrices at the site $m$ and $g$ plays the role of a (dimensionless) magnetic field that favors the alignment of the spins along the $x$-axis. This system exhibits a quantum phase transition between a paramagnetic phase ($|g|\gg1$) and a doubly degenerated phase with ferromagnetic order ($|g|\ll1$). There are two critical points at $g_{c}=\pm1$. We shall consider the nonadiabatic crossing of the critical point $g_{c}=-1$ after initializing the system in the paramagnetic phase and ending in a ferromagnetic phase. Our quantum simulation approach relies on the equivalence of in the TFQIM  in one spatial dimension and an ensemble of independent two-level systems. This seminal result forms the basis of much of the progress on the study of quantum phase transitions and can be established via the Jordan-Wigner transformation~\cite{JW28}, Fourier transform and Bogoliubov transformation~\cite{barouch1971pra}. We detail these steps in the Supplementary Note 1, where it is shown that the TFQIM Hamiltonian can be alternatively written in terms of independent modes as~\cite{Sachdev,Chakrabarti96} 
\begin{eqnarray}
\hat{\mathcal{H}}=\sum_{k>0}\hat{H}_{k}=\sum_{k>0}\epsilon_{k}\pap{g}\pap{\hat{\gamma}_{k}^{\dagger}\hat{\gamma}_{k}+\hat{\gamma}_{-k}^{\dagger}\hat{\gamma}_{-k}-1},
\end{eqnarray}
where $\hat{\gamma}_{k}$ are quasiparticle operators, with $k$ labeling each mode and taking values $k=\frac{\pi}{N}\pap{2m-1}$ with $m=-\frac{N}{2}+1,\ldots,\frac{N}{2}$. The energy $\epsilon_k$ of the $k$-th mode is $\epsilon_{k}\pap{g}=2J\sqrt{\pap{g-\cos k}^2 +\sin^2 k}$. Other  physical observables can as well be expressed in both the spin and momentum representations. We are interested in characterizing the number of kinks. As conservation of momentum  dictates that kinks are formed in pairs,  we focus on the kink-pair number operator $\hat{\mathcal{N}}~\equiv~\sum_{m=1}^{N}\left(\hat{1}-\hat{\sigma}_{m}^{z}\hat{\sigma}_{m+1}^{z}\right)/4$. The latter  can be equivalently written as $\hat{\mathcal{N}}=\sum_{k>0}\hat{\gamma}_{k}^{\dagger}\hat{\gamma}_{k}$, where $\hat{\gamma}_{k}^{\dagger}\hat{\gamma}_{k}$ is a Fermion number operator, with eigenvalues $\{0,1\}$. As different $k$-modes are decoupled,  this representation paves the way to simulate the dynamics of the phase transition in the TFQIM in ``momentum space'': the dynamics of each mode can be simulated with an ion-trap qubit, in which the  expectation value of $\hat{\gamma}_{k}^{\dagger}\hat{\gamma}_{k}$ can be measured. To this end, we consider the quantum critical dynamics induced by a ramp of the magnetic field
\begin{eqnarray}
g(t)=\frac{t}{\tau_{Q}}+g(0),\label{gt}
\end{eqnarray}
in a time scale $\tau_{Q}$ that we shall refer to as the quench time. We further choose $g(0)<-1$ in the paramagnetic phase. In momentum space, driving the phase transition is equivalently described by an ensemble of Landau-Zener crossings. This observation proved key in establishing the validity of the KZM in the quantum domain~\cite{Dziarmaga05,Xu2014,Cui16}: the average  number of kink pairs $\langle\hat{\mathcal{N}}\rangle=\langle n\rangle$ after the quench scales as 
\begin{eqnarray}
\langle n\rangle_{{\rm KZM}}=\frac{N}{4\pi}\sqrt{\frac{\hbar}{2J\tau_{Q}}},\label{navkzm}
\end{eqnarray}
in agreement with the universal power law $\langle n\rangle_{{\rm KZM}}\propto\tau_{Q}^{-\frac{\nu}{1+z\nu}}$ with critical exponents $\nu=z=1$ and one spatial dimension ($D=1$).
\section*{Results}
The full counting statistics of topological defects is encoded in the probability $P(n)$ that a given number of kink pairs $n$ is obtained as a measurement outcome of the observable $\hat{\mathcal{N}}$. To explore the implications of the scaling theory of phase transitions, we focus on the characterization of the probability distribution $P(n)$ of the  kink-pair number in the final nonequilibrium state upon completion of the crossing of the critical point induced by~\eqref{gt}. Exploiting the equivalence
between the spin and momentum representation, the dynamics in each mode leads to two possible outcomes, corresponding to the mode being found in the excited state (e) or the ground state (g), with probabilities $p_{e}=p_{k}$ and $p_{g}=1-p_{k}$, respectively. Thus, one can associate with each mode $k>0$ a discrete random variable, with excitation probability $p_{k}=\langle\hat{\gamma}_{k}^{\dagger}\hat{\gamma}_{k}\rangle$. The excitation probability of each mode is that of Bernoulli type. As the kink-pair number $n$ is associated with the number of modes excited, $P(n)$ is a Poisson binomial distribution, with the characteristic function~\cite{delcampo18}
\begin{eqnarray}
\widetilde{P}(\theta) =  \int_{-\pi}^{\pi}d\theta P(n)e^{in\theta}=\prod_{k>0}\left[1+\pap{e^{i\theta}-1} p_{k}\right],\label{charfunc}
\end{eqnarray}
associated with the sum of $N/2$ independent Bernoulli variables. The mean and variance of $P(n)$ are thus set by the respective sums of the mean and variance of the $N/2$ Bernoulli distributions characterizing each mode, $\langle n\rangle=\sum_{k>0}p_{k}$ and ${\rm Var}(n)=\sum_{k>0}p_{k}(1-p_{k})$. The full distribution $P(n)$ can be obtained via an inverse Fourier transform. The theoretical prediction of the kink distribution $P(n)$ relies on knowledge of the excitation probabilities $\{p_{k}\}$, which can be estimated using the Landau-Zener formula $p_{k}=\exp\left(-\frac{\pi}{\hbar}J\tau_{Q}k^{2}\right)$~\cite{Dziarmaga05} (details are given in Supplementary Note 2).
\begin{figure}[h!]
\centering \includegraphics[width=15cm]{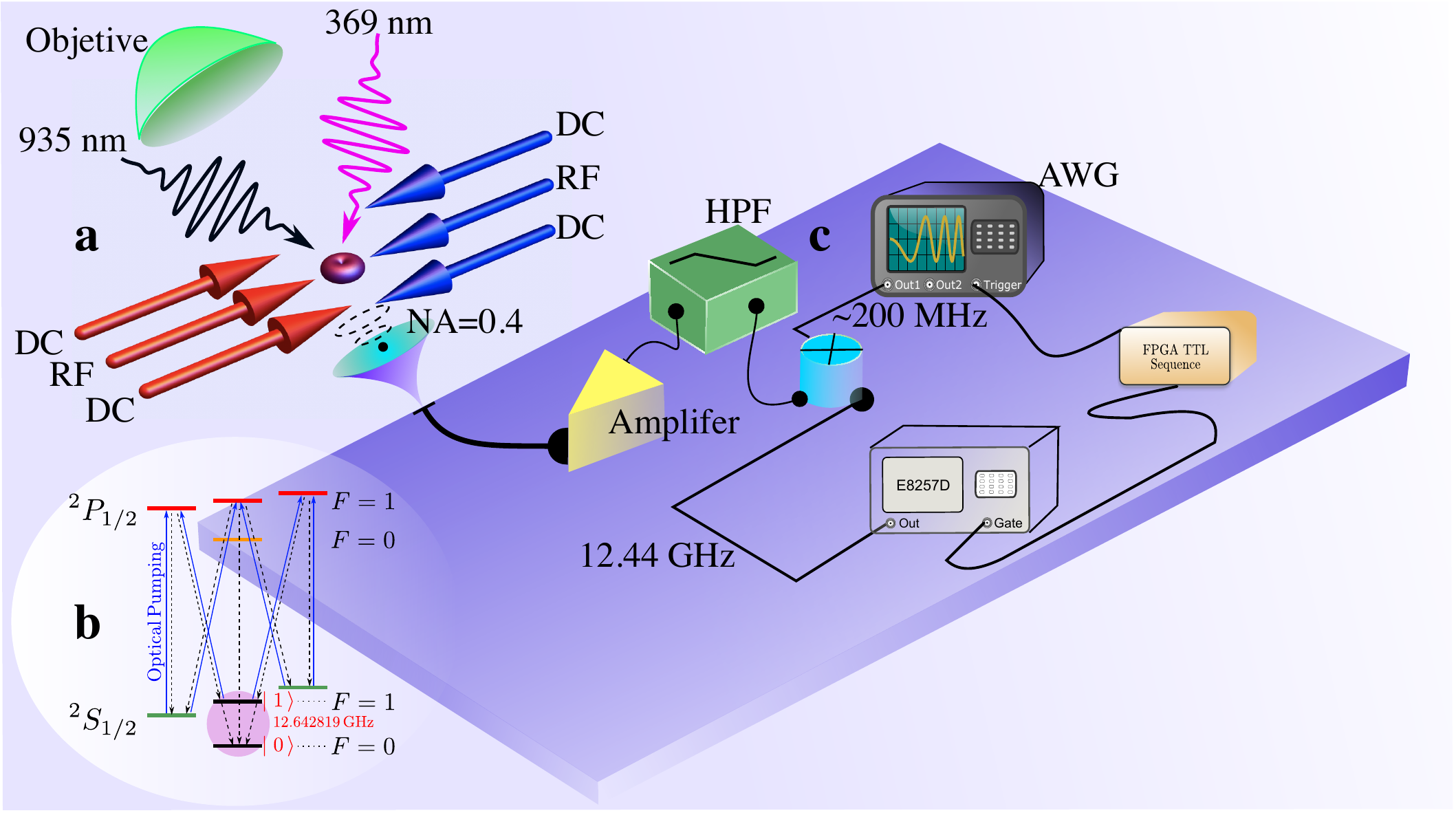} 
\caption{\label{fig_1}Experiment setup. \textbf{a} A single $^{171}{\mathrm{Yb}}^{+}$ is trapped in a needle trap, which consists of six needles on two perpendicular planes. \textbf{b}  The qubit energy levels are denoted by $\left|0\right\rangle$ and $\left|1\right\rangle $, which is the hyperfine clock transition of the trapped ion. \textbf{c} The qubit is driven by a microwave field, generated by a mixing wave scheme in the high pass filter (HPF).  Operations on the qubit are implemented by programming the arbitrary wave generator (AWG). The quantum critical dynamics of the one-dimensional TFQIM is detected by measuring corresponding Landau-Zener crossings governing the dynamics in each mode.}
\end{figure}

{\bf Experimental design.} Experimentally, the excitation probabilities $\{p_{k}\}$ can be measured by probing the dynamics in each mode, that is simulated with the ion-trap qubit. The dynamics of a single mode $k$ is described by a Landau-Zener crossing, that  is implemented with a $^{171}\mathrm{Yb^{+}}$ ion confined in a Paul trap consisting of six needles placed on two perpendicular planes, as shown in Fig.~\ref{fig_1}{\bf a}. The hyperfine clock transition in the ground state $S_{1/2}$ manifold is chosen to realize the qubit, with energy levels denoted by $\left|0\right\rangle \equiv\left|F=0,\,m_{F}=0\right\rangle $ and $\left|1\right\rangle \equiv\left|F=1,\,m_{F}=0\right\rangle $, as shown in Fig.~\ref{fig_1}{\bf b}.

{\it State preparation and manipulation:} At zero static magnetic field, the splitting between $\left|0\right\rangle $ and $\left|1\right\rangle $ is 12.642812 GHz. We applied a static magnetic field of 4.66 G to define the quantization axis, which changes the $\left|0\right\rangle $ to $\left|1\right\rangle $ resonance frequency to 12.642819 GHz, and creates a 6.5 MHz Zeeman splittings for $\mathrm{^{2}S_{1/2},\,F=1}$. In order to manipulate the hyperfine qubit with high control, coherent driving is implemented by a wave mixing method, see the scheme in Fig.~\ref{fig_1}{\bf c}. First, an arbitrary wave generator (AWG) is programmed to generate signals around 200 MHz. Then, the waveform is mixed with a 12.442819 GHz microwave (generated by Agilent, E8257D) by a frequency mixer. After the mixing process, there will be two waves around 12.242 GHz and 12.642 GHz, so a high pass filter is used to remove the 12.224 GHz wave. Finally, the wave around 12.642 GHz is amplified to 2\,W  and used to irradiate the trapped ion with a horn antenna.

{\it Measurement:} For a typical experimental measurement in a single mode, Doppler cooling is first applied to cool down the ion~\cite{Olmschenk2007}. The ion qubit is then initialized in the $\left|0\right\rangle $ state, by applying a resonant light at 369nm to excite ${S_{1/2}},\,F=1$ to ${P_{1/2}},\,F=1$. Subsequently, the programmed microwave is started to drive the ion qubit. Finally, the population of the bright state $\left|1\right\rangle $ is detected by fluorescence detection with another resonant light at 369nm, exciting ${S_{1/2}},\,F=1$ to ${P_{1/2}},\,F=0$. Fluorescence of the ion is collected by an objective with 0.4 numerical aperture (NA). A 935nm laser is used to prevent the state of the ion to jump to metastable states \cite{Olmschenk2007}.  The initialization process can prepare the $\left|0\right\rangle $ state with fidelity $>99.9\%$. The total error associated with the state preparation and measurement is measured as 0.5\% ~\cite{Maunz2017NC}.

\begin{figure}[h!]
\centering \includegraphics[width=9cm]{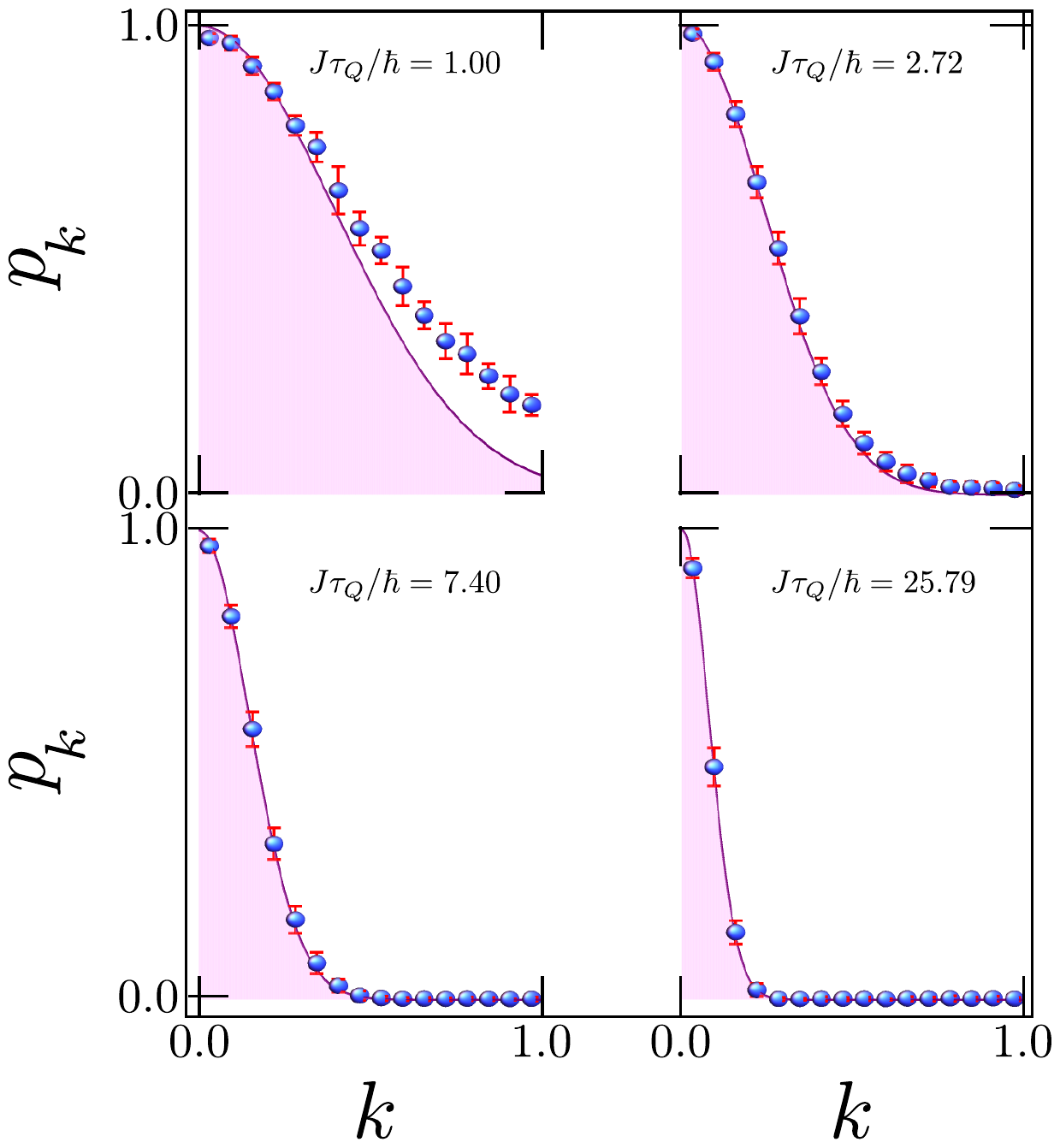} \caption{\label{fig_2}\textbf{Excitation probability in the ensemble of Landau-Zener crossings.} The quantum critical dynamics of the one-dimensional TFQIM is accessible in an ion-trap quantum simulator, which implements the corresponding Landau-Zener crossings governing the dynamics in each mode. The Rabi frequency $\Omega_R=4J/\hbar$ was set to $2\pi\times 20$ kHz in experiment. For each value of $k$, the excitation probability is estimated from $10000$ measurements. The shaded region describes the excitation probability predicted by  the Landau-Zener formula. Deviations from the latter become apparent for fast quench times, especially for  large values of $k$.}
\end{figure}

In the Supplementary Note 1, we rewrite the explicit Hamiltonian of the $k$-th mode as a linear combination of a single two-levels system~\cite{Dziarmaga05}. In this way, experimentally, the Hamiltonian of a single $k$-mode can be explicitly mapped to a qubit Hamiltonian describing a two-level  system driven by a chirped  microwave pulse, 
\begin{equation}
\hat{H}_{k}^{\rm TLS} =\frac{1}{2}\hbar\pap{\Delta_k(t)\hat{\sigma}_{z}+\Omega_R\hat{\sigma}_{x}}, 
\end{equation}
where $\Omega_R=4J/\hbar$ and  $\Delta_k(t)=4J[g(t)+\cos{k}]/(\hbar\sin{k})$ are the Rabi frequency and the detuning of the chirped pulse, respectively. 
To measure the excitations probability after a finite ramp with normalized quench parameter $A=J\tau_Q/\hbar$, we can vary $g(t)$ from -5 to 0. For this purpose, a chirped pulse with length of  $T_p=5A T_R\sin{k}$ and $g(t)=5t/T_p-5$ is used in the experiment, with $T_R=1/\Omega_R$ denoting the Rabi time. A typical operation process driven by the programmed microwave is illustrated on the Bloch sphere, see Supplementary Notes 3 and 4 for details. First, the qubit is prepared in the ground state of $\hat{H}_{k}^{\rm TLS}$ at $g(0)=-5$. Then, $g(t)$ is ramped from $-5$ to 0 with quenching rate $1/\tau_{Q}$. Finally, the ground state of the final $\hat{H}_{k}^{\rm TLS}$ is rotated to qubit $\left|0\right>$, so that the excited state is mapped to $\left|1\right>$,  which can be detected as a bright state. The measured excitation probability for different quench times is shown in Fig.~\ref{fig_2}.

{\bf Full probability distribution of topological defects.} From the experimental data, the distribution $P(n)$ of the number of kink pairs is obtained using the characteristic function~\eqref{charfunc} of a Poisson binomial distribution in which the probability $p_{k}$ of each Bernoulli trial is set by the experimental value of the excitation probability upon completion of the corresponding Landau-Zener sweep. This result is compared with the theoretical prediction valid in the scaling limit --the regime of validity of the KZM-- in which $P(n)$ approaches the normal distribution, away from the adiabatic limit ~\cite{delcampo18} 

\begin{eqnarray}
P(n)\simeq\frac{1}{\sqrt{6\langle n\rangle_{{\rm KZM}}/\pi}}\exp\left[-\frac{\pi^{2}(n-\langle n\rangle_{{\rm KZM}})^{2}}{6\langle n\rangle_{{\rm KZM}}}\right].\label{Gausspn}
\end{eqnarray}

A comparison between $P(n)$ and Equation~\eqref{Gausspn} is shown in Fig.~\ref{fig_3}. The matching between theory and experiment is optimal in the scaling limit far away from the onset of the adiabatic dynamics or fast quenches, for which non-universal corrections are expected. We further note that the onset of adiabatic dynamics enhances non-normal features of the experimental $P(n)$ that cannot be simply accounted for by a truncated Gaussian distribution, that takes into account the fact that the number of kinks $n=0,1,2,3,\dots$
\begin{figure}[h!]
\centering \includegraphics[width=14cm]{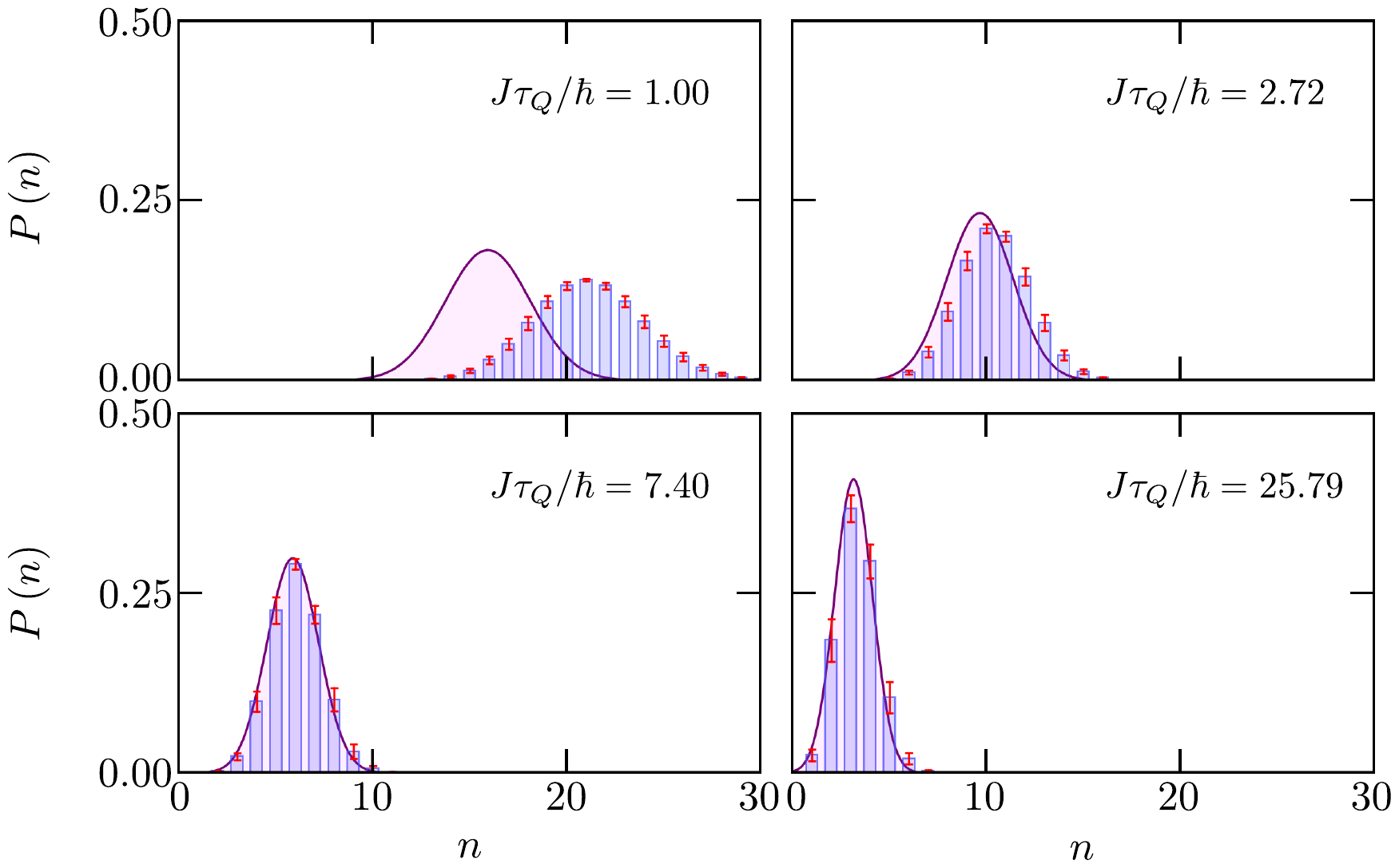} \caption{\label{fig_3}\textbf{Probability distribution of the number of kink pairs $P(n)$ generated as a function of the quench time.} The experimental kink-pair number distribution for a TFQIM with $N=100$ spins is compared with the Gaussian approximation derived in the scaling limit, Eq.~\eqref{Gausspn}, ignoring high-order cumulants. The mean and width of the distribution are reduced as the quench time is increased. The experimental $P(n)$ is always broader and shifted to higher kink-pair numbers than the theoretical prediction. Non-normal features of $P(n)$ are enhanced near the sudden-quench limit and at the onset of adiabatic dynamics.}
\end{figure}

\begin{figure}[t!]
\centering \includegraphics[width=8.5cm]{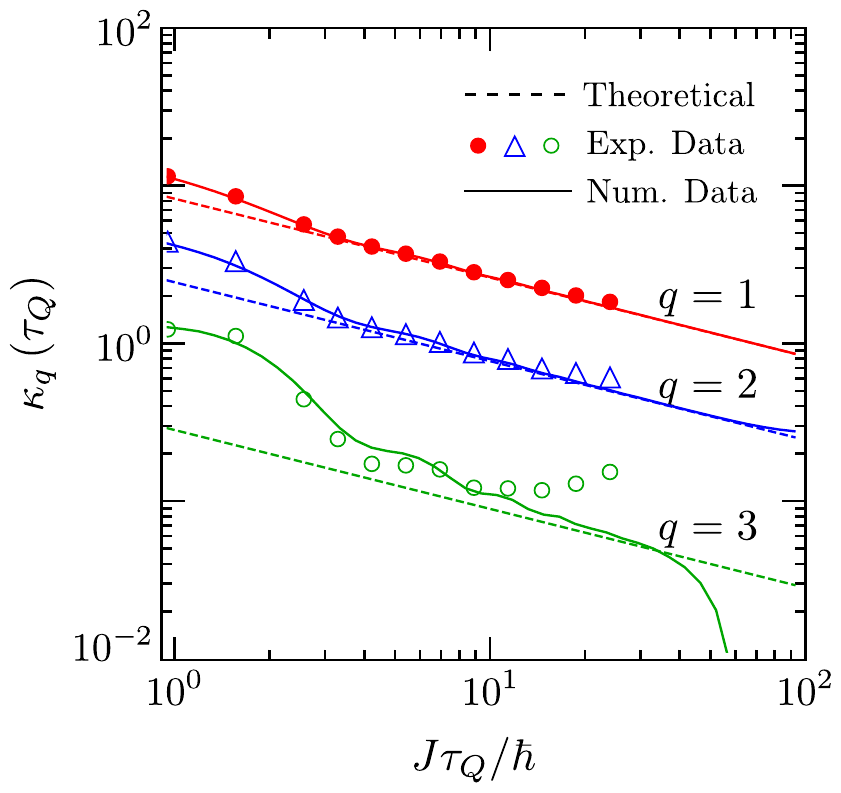} \caption{\label{fig_4}\textbf{Universal scaling of the cumulants $\kappa_{q}(\tau_{Q})$ of the kink-pair number distribution $P(n)$ as a function of the quench time.} The experimental data (symbols) is compared with the scaling prediction (dashed lines) and the numerical data (solid lines) for the first three cumulants with $q=1,2,3$. The universal scaling of the first cumulant $\kappa_{1}(\tau_{Q})=\langle n\rangle$ is predicted
by the KZM according to Eq.\eqref{navkzm}. Higher-order cumulants are also predicted to exhibit a universal scaling with the quench time. All cumulants of the experimental $P(n)$ exhibit deviations from the universal scaling at long quench times consistent with a dephasing-induced anti-KZM behavior. Further deviations from the scaling behavior are observed at fast quench times. The range of quench times characterized by universal behavior is reduced for high-order cumulants as $q$ increases. The error bars are much smaller than the size of the symbols used to depict the measured points.}
\end{figure}

To explore universal features in $P(n)$, we shall be concerned with the scaling of its cumulants $\kappa_{q}$ ($q=1,2,3\dots$) as a function of the quench time. We focus on the first three, theoretically derived in the scaling limit in~\cite{SM}. The first one is set by the KZM estimate for the mean  number of kink pairs $\kappa_{1}(\tau_{Q})=\langle n\rangle_{{\rm KZM}}$. The second one equals the variance and as shown in the SM is set by $\kappa_{2}(\tau_{Q})={\rm Var}(n)=(1-1/\sqrt{2})\langle n\rangle_{{\rm KZM}}$. Finally, the third cumulant is given by the third centered moment $\kappa_{3}(\tau_{Q})=\langle(n-\langle n\rangle)^{3}\rangle=(1-3/\sqrt{2}+2/\sqrt{3})\langle n\rangle_{{\rm KZM}}$. Indeed, all cumulants are predicted to be nonzero and proportional to $\langle n\rangle_{{\rm KZM}}$~\cite{delcampo18}, with  the scaling of the first cumulant $\langle n\rangle_{{\rm KZM}}\propto\tau_{Q}^{-\frac{1}{2}}$ being dictated by the KZM.  We compared this theoretical prediction with the cumulants of the experimental
$P(n)$ in Fig.~\ref{fig_4}, represented by dashed lines and symbols, respectively.  As discussed in the Supplementary Information, deviations from the power law occur at very fast quench times, satisfying $\tau_{Q}<\hbar/(\pi^{3}J)$. However,  this regime  is not probed in the experiment, as all data points are  taken for larger values of $\tau_Q$. Within the parameters explored,  deviations at fast quenches with $J\tau_{Q}/\hbar\sim1$ are due to the fact that the excitation probability $p_{k}$ in each mode is not accurately described by the Landau-Zener formula, as shown in Fig. \ref{fig_2}. 
For moderate ramps, the power-law scaling of the cumulants is verified for $q=1,2$. 
The power-law scaling for $\kappa_{q}(\tau_{Q})$ with $q>1$ establishes the universal character of critical dynamics beyond the KZM. The latter is explicitly verified for $\kappa_{2}(\tau_{Q})$, as shown in Fig.~\ref{fig_4}. Beyond the fast-quench deviations shared by the first cumulant, the power law scaling of the variance of  the number of kink pairs  $\kappa_{2}(\tau_{Q})$ extends to all the larger values of the quench time explored, with barely noticeable deviations.
By contrast, for  the third cumulant $\kappa_{3}(\tau_{Q})$ the experimental data is already dominated by non-universal contributions both at short quench times, away from the scaling limit. For slow ramps, the third cumulant exhibits an onset of adiabatic dynamics due to finite size effects around $J\tau_Q/\hbar=10^2$. Finite-size effects are predicted to lead to a sharp suppression of the third cumulant. However, the experimental data shows that $\kappa_3$ starts to increase with   $\tau_Q$. This is reminiscent of the anti-Kibble-Zurek behavior that has been reported in the literature for the first cumulant in the presence of heating sources~\cite{Griffin12,Dutta16} and we attribute it to  
 noise-induced dephasing of the trapped-ion qubit. The nature of these deviations is significantly more pronounced in high-order cumulants,  reducing the regime of applicability of the universal scaling. 
 
\section*{Discussion}
In summary, using a trapped-ion quantum simulator we have probed the full counting statistics of topological defects in the quantum Ising chain, the paradigmatic model of quantum phase transitions. The statistics of the number of kink pairs has been shown to be described by the Poisson binomial distribution, with cumulants obeying a universal power-law with the quench time in which the phase transition is crossed. Our findings demonstrated that the scaling theory associated with a critical point rules the formation of topological defects beyond the scope of the Kibble-Zurek mechanism, which is restricted to the average number. Our work could
be extended to probe systems with topological order in which defect formation has been predicted to be anomalous~\cite{Bermudez09}. We anticipate that the universal features of the full counting statistics of topological defects may be used in the error analysis of adiabatic quantum annealers, where the Kibble-Zurek mechanism already provides useful heuristics~\cite{Suzuki09b}.

\section*{Methods}
A detailed description of analytical methods and experimental techniques employed to obtain the results can be found in the Supplementary Information accompanying this work.
\section*{Data availability}
The data that support the plots and other findings within this paper are available from the corresponding author upon reasonable request.

\bibliography{FCSKinkExp_Bib_v2}{}
\bibliographystyle{apsrev4-1}
\section*{Acknowledgements}
We acknowledge funding support from the National Key Research and Development Program of China (Nos. 2017YFA0304100, 2016YFA0302700), the National Natural Science Foundation of China (Nos. 61327901, 11774335, 11474268, 11734015, 11821404), Key Research Program of Frontier Sciences, CAS (No. QYZDY-SSW-SLH003), the Fundamental Research Funds for the Central Universities (Nos. WK2470000026, WK2470000018), Anhui Initiative in Quantum Information Technologies (Nos. AHY020100, AHY070000), and John Templeton Foundation.
\section*{Author contributions}	
A.d.C. proposed the project and developed the theoretical framework. J.-M.C. and Y.-F.H. took the experimental measurements. F. J. G.-R. developed numerical simulations and prepared the figures. All authors contributed to the analysis of the results and the writing of the manuscript. C.-F.L., G.-C.G. and A.d.C. supervised the project.
\section*{Additional information}
Supplementary information accompanies this paper at url will be inserted by publisher.
\section*{Competing Interests}	
The authors declare that they have no competing financial interests. 
\section*{Correspondence}
Correspondence and requests for materials should be addressed to Y.-F. H.~(email: \href{mailto:hyf@ustc.edu.cn}{hyf@ustc.edu.cn}), C.-F. L.~(email: \href{cfli@ustc.edu.cn}{cfli@ustc.edu.cn}) and A. d. C.~(email: \href{mailto:adolfo.delcampo@dipc.org}{adolfo.delcampo@dipc.org}).	
\newpage
\pagebreak
\clearpage
\setcounter{equation}{0}
\setcounter{figure}{0}
\setcounter{table}{0}
\setcounter{section}{0}
\setcounter{page}{1}
\makeatletter
\renewcommand{\thefigure}{\arabic{figure}}
\renewcommand{\figurename}{{\bf Supplementary Figure.}}
\renewcommand{\theequation}{S\arabic{equation}}
\renewcommand{\thesection}{{\large {\bf Supplementary Note \arabic{section}}}}
\renewcommand{\thesubsection}{\arabic{subsection}}
\renewcommand{\bibsection}{\section*{Supplementary Notes: References}}

\newcounter{fnnumber}
\renewcommand{\thefootnote}{\fnsymbol{footnote}}
\begin{center}
\textbf{\large ---Supplementary Notes---\\
Experimentally testing quantum critical dynamics beyond the Kibble-Zurek mechanism}\\
\vspace{0.5cm}
Jin-Ming Cui$^{1,2}$, Fernando Javier G{\'o}mez-Ruiz$^{3,4,6}$, Yun-Feng Huang$^{1,2,*}$,\\
Chuan-Feng Li$^{1,2,\dagger}$, Guang-Can Guo$^{1,2}$ \& Adolfo del Campo$^{3,5,6,7,\ddagger}$\\
\vspace{0.2cm}
$^{1}${\it CAS Key Laboratory of Quantum Information, University of Science and Technology of China, Hefei 230026, China}\\
$^{2}${\it CAS Center For Excellence in Quantum Information and Quantum Physics, University of Science and Technology of China, Hefei 230026, China}\\
$^{3}${\it Donostia International Physics Center, E-20018 San Sebasti{\'a}n, Spain}\\
$^{4}${\it Departamento de F{\'i}sica, Universidad de los Andes, A.A. 4976, Bogot{\'a} D. C., Colombia}\\
$^{5}${\it  IKERBASQUE, Basque Foundation for Science, E-48013 Bilbao, Spain}\\
$^{6}${\it  Department of Physics, University of Massachusetts Boston, 100 Morrissey Boulevard, Boston, MA 02125}\\
$^{7}${\it Theoretical Division, Los Alamos National Laboratory, MS-B213, Los Alamos, NM 87545, USA}
\end{center}
\textbf{This PDF file includes:} 
\begin{itemize}
	\item \textbf{Supplementary Note 1.} Mapping between spin and free-Fermion
	representations in the Ising model. 
	\item \textbf{Supplementary Note 2.} Cumulants of the kink-pair number distribution
	$P(n)$. 
	\item \textbf{Supplementary Note 3.} Experimental Setup. 
	\item \textbf{Supplementary Figure 1.} Universal scaling of the cumulants
	$\kappa_{q}(\tau_{Q})$ of the kink-pair number distribution $P(n)$
	as a function of the quench time. 
	\item \textbf{Supplementary Figure 2.} Scheme to measure the excitation
	probability. 
	\item \textbf{Supplementary Figure 3.} Histogram for photon counts in state
	preparation and detection experiments. 
\end{itemize}

\section{Mapping between spin and free-Fermion representations in the Ising
	model}

We consider the one-dimensional quantum Ising model in a transverse
magnetic field $g$. The TFQIM Hamiltonian is given by 
\begin{equation}
\hat{\mathcal{H}}=-J\sum_{m=1}^{N}(\hat{\sigma}_{m}^{z}\hat{\sigma}_{m+1}^{z}+g\hat{\sigma}_{m}^{x}),\label{H1}
\end{equation}
where $J$ is the hopping parameter and $\hat{\sigma}_{m}^{\alpha}$
are the Pauli matrix along the direction $\alpha=x,z$  at site $m$. Here
we  assume for convenience that $N$ is even and impose periodic boundary conditions
$\hat{\sigma}_{N+1}^{z}=\hat{\sigma}_{1}^{z}$. The Hamiltonian is
mapped to a free fermion model through the Jordan-Wigner transformation~\cite{SM_barouch1971pra}
\begin{align}
\hat{\sigma}_{m}^{z} & =-\pap{\hat{c}_{m}^{\dagger}+\hat{c}_{m}}\prod_{m'<m}\pap{1-2\hat{c}_{m'}^{\dagger}\hat{c}_{m'}}, & \hat{\sigma}_{m}^{x} & =1-2\hat{c}_{m}^{\dagger}\hat{c}_{m}.\label{JWT}
\end{align}
Inserting ~\eqref{JWT} into the Ising Hamiltonian Eq.~\eqref{H1} 
we find that the Hamiltonian is quadratic in the Fermi operators 
\begin{equation}
\begin{split}\hat{\mathcal{H}}= & -J\left[\sum_{m=1}^{N-1}\hat{c}_{m}^{\dagger}\hat{c}_{m+1}+\hat{c}_{m+1}^{\dagger}\hat{c}_{m}+\hat{c}_{m}\hat{c}_{m+1}+\hat{c}_{m}^{\dagger}\hat{c}_{m+1}^{\dagger}+g\sum_{m=1}^{N}\pap{1-2\hat{c}_{m}^{\dagger}\hat{c}_{m}}\right.\\
& \left.+\hat{\Pi}\pap{\hat{c}_{N}^{\dagger}\hat{c}_{1}+\hat{c}_{N}\hat{c}_{1}+\hat{c}_{1}^{\dagger}\hat{c}_{N}+\hat{c}_{1}^{\dagger}\hat{c}_{N}^{\dagger}}\right].
\end{split}
\end{equation}
This Hamiltonian includes terms such as $\hat{c}_{m}\hat{c}_{m+1}$
or $\hat{c}_{m+1}^{\dagger}\hat{c}_{m}^{\dagger}$ that violate the
fermion-number conservation. However, the parity operator $\hat{\Pi}=\pap{-1}^{\sum_{m=1}^{N}\hat{c}_{m}^{\dagger}\hat{c}_{m}}$
is a constant of motion having the value $+1$ and $-1$. We expand
$\hat{c}_{m}$ into the Fourier modes 
\begin{equation}
\hat{c}_{m}=\frac{e^{-i\pi/4}}{\sqrt{N}}\sum_{k}\exp\pas{im\pap{ka}}\hat{c}_{k},
\end{equation}
where the pseudomomenta $k$ take values 
\begin{equation}
k=\frac{\pi}{Na}\pap{2j-1},\qquad j=-\frac{L}{2}+1,\ldots,\frac{L}{2}.
\end{equation}
Using the choice of even $N$ and considering anti-periodic boundary
condition $\hat{c}_{N+1}=-\hat{c}_{N}$, the Hamiltonian maps to 
\begin{equation}
\hat{\mathcal{H}}_{{\rm Even}}=-\sum_{0<k<\frac{\pi}{a}}\begin{pmatrix}\hat{c}_{k}^{\dagger} & \hat{c}_{-k}\end{pmatrix}\hat{H}_{k}\begin{pmatrix}\hat{c}_{k}\\
\hat{c}_{-k}^{\dagger}
\end{pmatrix},
\end{equation}
where $\hat{H}_{k}$ is defined by 
\begin{equation}
\hat{H}_{k}\equiv2J\pap{g-\cos\pap{ak}}\hat{\sigma}^{z}+2J\sin\pap{ak}\hat{\sigma}^{x}.\label{H3K}
\end{equation}
Next, we use the Bogoliubov transformation to rewrite the Hamiltonian in terms of a new set of
fermionic operators $\hat{\gamma}_{k}$ whose number is conserved.
These new operators are defined via a unitary transformation in terms of the
pair $\hat{c}_{k}$, $\hat{c}_{-k}^{\dagger}$: 
\begin{align*}
\hat{\gamma}_{k}=u_{k}\hat{c}_{k}+v_{k}\hat{c}_{-k}^{\dagger},
\end{align*}
where $u_{k}$, $v_{k}$ are numbers satisfying $u_{k}^{2}+v_{k}^{2}=1$,
$u_{-k}=u_{k}$, and $v_{-k}=-v_{k}$. It can be checked that canonical
fermion anti-commutation relation for the $\hat{c}_{k}$ imply that
the same relations are also satisfied by $\hat{\gamma}_{k}$, that
is, 
\[
\left\{ \hat{\gamma}_{k},\hat{\gamma}_{k'}^{\dagger}\right\} =\delta_{k,k'},\qquad\left\{ \hat{\gamma}_{k}^{\dagger},\hat{\gamma}_{k'}^{\dagger}\right\} =\left\{ \hat{\gamma}_{k},\hat{\gamma}_{k'}\right\} =0.
\]
The Hamiltonian is diagonalized in terms of $\hat{\gamma}_{k}$, 
\begin{equation}
\hat{\mathcal{H}}_{{\rm Even}}=\sum_{0<k<\frac{\pi}{a}}\epsilon_{k}\pap{g}\pap{\hat{\gamma}_{k}^{\dagger}\hat{\gamma}_{k}+\hat{\gamma}_{-k}^{\dagger}\hat{\gamma}_{-k}-1},
\end{equation}
where $\epsilon_{k}\left(g\right)=2J\sqrt{\left(g-\cos\left(ak\right)\right)^{2}+\sin^{2}\left(ak\right)}$,
with the ground state defined $\hat{\gamma}_{k}\left|\textbf{0}\right\rangle =0$.
The ground state free energy reads 
\begin{eqnarray*}
	\epsilon_{0}\equiv\frac{E_{0}}{NJ}=-\frac{1}{2\pi}\int_{-\pi}^{\pi}\epsilon_{k}\pap{g}dk.
\end{eqnarray*}

\section{Cumulants of the kink-pair number distribution $P(n)$}

The kink-pair number distribution $P(n)$ describes the probability
of observing $n$ pairs of kinks. The number of kink pairs is measured
by the observable 
\begin{eqnarray}
\hat{\mathcal{N}}=\frac{1}{4}\sum_{m=1}^{N}(1-\hat{\sigma}_{m}^{z}\hat{\sigma}_{m+1}^{z})=\sum_{k\geq0}\hat{\gamma}_{k}^{\dagger}\hat{\gamma}_{k},
\end{eqnarray}
with eigenvalues $n=0,1,2,\dots$ The kink-pair number distribution
is thus defined as the quantum expectation value of the projector
operator $\delta[\hat{\mathcal{N}}-n]$ on the subspace with $n$
pairs of kinks 
\begin{eqnarray}
P(n)={\rm tr}\pas{\hat{\rho}\,\delta[\hat{\mathcal{N}}-n]},
\end{eqnarray}
where the expectation value is taken over the final state $\hat{\rho}$
upon completion of the quench. The Fourier transform of $P(n)$ is
the characteristic function 
\begin{eqnarray}
\widetilde{P}(\theta) & = & \int_{-\pi}^{\pi}d\theta P(n)e^{in\theta}=\mathbb{E}\left[e^{in\theta}\right].\label{charfunc}
\end{eqnarray}
The cumulants $\kappa_{q}$ of the distribution $P(n)$ are defined
via the expansion of the cumulant-generating function, which is the logarithm
of the characteristic function, 
\begin{eqnarray}
\log\widetilde{P}(\theta)=\sum_{q=1}^{\infty}\frac{(i\theta)^{q}}{q!}\kappa_{q}.
\end{eqnarray}

We next focus on the characterization of $P(n)$ for the state that
results from the nonadiabatic crossing of the phase transition in
the TFQIM and derive the exact expression for
the first few cumulants $\{\kappa_{q}\}$. To this end, we note that
the nonadiabatic crossing of the phase transition results in the production
of quasi-particles (and kinks) in pairs. Specifically, due to the
conservation of momentum, quasiparticles with wavectors $+k$ and
$-k$ are excited jointly. The distribution of the number of pairs
of kinks is a Poisson binomial distribution associated with $N/2$
Bernoulli trials each with probability $p_{k}$ for $k>0$. Its characteristic
function reads 
\begin{eqnarray}
\widetilde{P}(\theta)=\prod_{k>0}\left[1+\pap{e^{i\theta}-1}p_{k}\right],
\end{eqnarray}
with $k\in\{\pi/N,3\pi/N,\ldots,(N-1)\pi/N\}$ for an Ising chain with
period boundary conditions and $p_{k}$ denoting the excitation probability
in the $k$-mode. The Poisson binomial distribution is well studied and
its cumulants are known to be of the form 
\begin{eqnarray}
\kappa_{1} & = & \langle n\rangle=\sum_{k>0}p_{k},\\
\kappa_{2} & = & {\rm Var}(n)\rangle=\sum_{k>0}p_{k}(1-p_{k}),\\
\kappa_{3} & = & \langle(n-\langle n\rangle)^{3}\rangle=\sum_{k>0}p_{k}(1-p_{k})(1-2p_{k}).
\end{eqnarray}

Their explicit evaluation can be performed using the Landau-Zener
formula for $p_{k}$. In the continuum limit, this yields 
\begin{align*}
\kappa_{1}\equiv\langle n\rangle & =\sum_{k>0}p_{k}\\
& =\frac{N}{2\pi}\int_{0}^{\pi}dk\exp\left(-\frac{1}{\hbar}2\pi J\tau_{Q}k^{2}\right).
\end{align*}
In terms of the error function, it is found that 
\begin{eqnarray}
\kappa_{1}=\frac{N}{4\pi}\sqrt{\frac{\hbar}{2J\tau_{Q}}}\times{\rm erf}\left(\pi\sqrt{\frac{2\pi J\tau_{Q}}{\hbar}}\right),\label{k1c}
\end{eqnarray}
an expression that is valid even for moderately fast quenches, as
long as described by the Landau-Zener formula but possibly away from
the scaling limit. However, for large $\tau_{Q}>\hbar/(2\pi^{3}J)$,
${\rm erf}\left(\frac{\sqrt{\pi}}{2d}\right)$ approaches unity and
the average number of kink pair simply reads 
\begin{eqnarray}
\langle n\rangle_{{\rm KZM}}=\frac{N}{4\pi}\sqrt{\frac{\hbar}{2J\tau_{Q}}},\label{k1a}
\end{eqnarray}
in agreement with the KZM~\cite{SM_Dziarmaga05}.

A similar derivation shows that the variance is given by 
\begin{align*}
\kappa_{2} & \equiv{\rm Var}(n)=\sum_{k>0}p_{k}(1-p_{k})\\
& =\frac{N}{2\pi}\int_{0}^{\pi}dke^{-\frac{2\pi J\tau_{Q}}{\hbar}k^{2}}\left(1-e^{-\frac{2\pi J\tau_{Q}}{\hbar}k^{2}}\right).
\end{align*}
This results in the exact expression 
\begin{eqnarray}
{\rm Var}(n)=\left[{\rm erf}\left(\sqrt{\frac{2\pi^{3}J\tau_{Q}}{\hbar}}\right)-\frac{1}{\sqrt{2}}{\rm erf}\left(\sqrt{\frac{4\pi^{3}J\tau_{Q}}{\hbar}}\right)\right]\langle n\rangle_{{\rm KZM}},\label{k2c}
\end{eqnarray}
that in the scaling limit reduces to 
\begin{eqnarray}
{\rm Var}(n)=\left(1-\frac{1}{\sqrt{2}}\right)\langle n\rangle_{{\rm KZM}}.\label{k2a}
\end{eqnarray}

To characterize the non-normal features of the distribution of kinks,
we also provide the exact expression for the third cumulant $\kappa_{3}$,
that equals the third centered moment $\mu_{3}=\langle(n-\langle n\rangle)^{3}\rangle$.
In particular, we note that $\kappa_{3}$ is related to the skewness
$\gamma_{1}$ of the distribution, as $\gamma_{1}=\kappa_{2}/\kappa_{2}^{3/2}$.
In the continuum, it reads 
\begin{eqnarray}
\kappa_{3} & \equiv & \langle(n-\langle n\rangle)^{3}\rangle=\sum_{k>0}p_{k}(1-p_{k})(1-2p_{k})\nonumber \\
& = & \frac{N}{2\pi}\int_{0}^{\pi}dke^{-\frac{2\pi J\tau_{Q}}{\hbar}k^{2}}\!\!\left(1-e^{-\frac{2\pi J\tau_{Q}}{\hbar}k^{2}}\right)\!\!\left(1-2e^{-\frac{2\pi J\tau_{Q}}{\hbar}k^{2}}\right)\nonumber \\
& = & \left[{\rm erf}\left(\sqrt{\frac{2\pi^{3}J\tau_{Q}}{\hbar}}\right)-\frac{3}{\sqrt{2}}{\rm erf}\left(\sqrt{\frac{4\pi^{3}J\tau_{Q}}{\hbar}}\right)+\frac{2}{\sqrt{3}}{\rm erf}\left(\sqrt{\frac{6\pi^{3}J\tau_{Q}}{\hbar}}\right)\right]\langle n\rangle_{{\rm KZM}}.\label{k3c}
\end{eqnarray}
This expression is simplified in the scaling limit, performing a power
series expansion with $\tau_{Q}\rightarrow\infty$. To leading order,
we obtain the expression quoted  in the main text 
\begin{eqnarray}
\kappa_{3}=\left(1-\frac{3}{\sqrt{2}}+\frac{2}{\sqrt{3}}\right)\langle n\rangle_{{\rm KZM}}.\label{k3a}
\end{eqnarray}

\begin{figure}[t]
	\begin{centering}
		\includegraphics[width=0.6\textwidth]{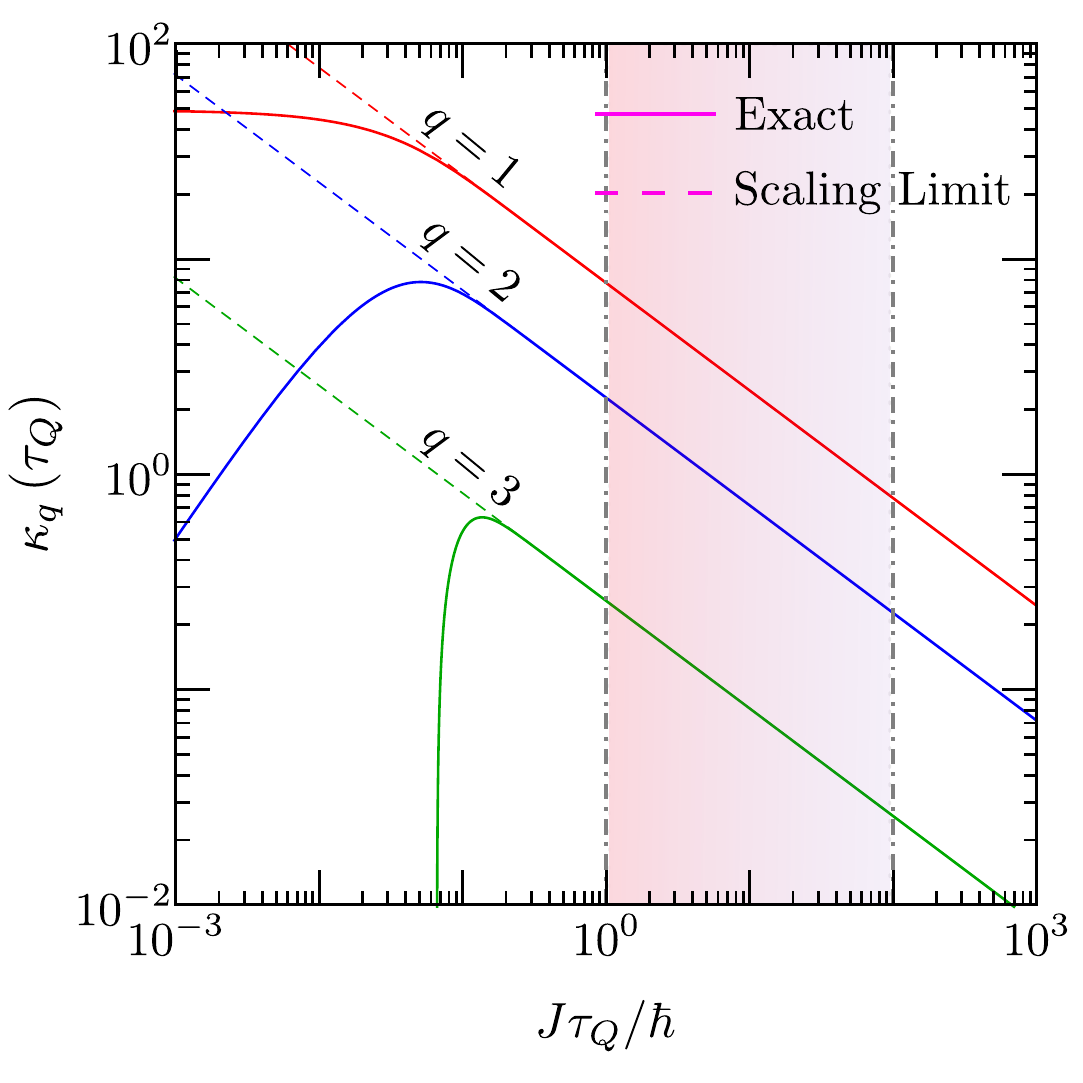} 
		\par\end{centering}
	\caption{\textbf{Universal scaling of the cumulants $\kappa_{q}(\tau_{Q})$
			of the kink-pair number distribution $P(n)$ as a function of the
			quench time.} The solid lines show the exact result for the first
		three cumulant $\kappa_{q}(\tau_{Q})$ with $q\in{1,2,3}$, as
		predicted by equations~\eqref{k1c},~\eqref{k2c}, and~\eqref{k3c}.
		The dashed lines show the universal scaling of the first cumulant
		according to Eqs.~\eqref{k1a},~\eqref{k2a}, and~\eqref{k3a}.
		In the experimental regime, marked by the shaded area, the scaling
		behavior holds.}
	\label{fig_3} 
\end{figure}

Note that the scaling expressions Eqs.~\eqref{k1a},~\eqref{k2a} and~\eqref{k3a} are obtained from the more general ones, Eqs.~\eqref{k1c},~\eqref{k2c} and~\eqref{k3c}, whenever the error functions can be
replaced by unity. The error function increases monotonically and
swiftly from zero value to unity as the argument is increased from
zero value. Indeed, it saturates at unity for fairly moderate values
of the argument, e.g., ${\rm erf}(2)=0.995$. By imposing $\sqrt{\frac{2q\pi^{3}J\tau_{Q}}{\hbar}}>2$,
we predict the onset of the scaling limit for $\kappa_{q}$ at 
\begin{eqnarray}
\tau_{Q}>\frac{2\hbar}{q\pi^{3}J}.
\end{eqnarray}
This condition is satisfied in all the experimental realizations we have performed. As a result, it is justified to consider the scaling
limit, in which it is actually possible to derived a closed form expression
of the characteristic function~\cite{SM_delcampo18} 
\begin{eqnarray}
\widetilde{P}(\theta)=\exp\left[-\langle n\rangle_{{\rm KZM}}\,{\rm Li}_{3/2}(1-e^{i\theta})\right],\label{CGFa}
\end{eqnarray}
in terms of the polylogarithmic function ${\rm Li}_{3/2}(x)=\sum_{p=1}^{\infty}x^{p}/p^{3/2}$.

From this expression, it follows that all cumulants are actually nonzero
and proportional to $\langle n\rangle_{{\rm KZM}}$. The kink-pair
number distribution is manifestly non-normal. The Gaussian (normal)
approximation discussed in the main text follows from neglecting all
$\kappa_{q}$ with $q>2$. One can also approximate $1-1/\sqrt{2}\approx3/\pi^{2}$,
so that 
\begin{eqnarray}
\widetilde{P}(\theta)\simeq\exp\left[-\langle n\rangle_{{\rm KZM}}\left(i\theta-\frac{3}{2\pi^{2}}\theta^{2}\right)\right],
\end{eqnarray}
and thus 
\begin{eqnarray}
P(n)\simeq\frac{1}{\sqrt{6\langle n\rangle_{{\rm KZM}}/\pi}}\exp\left[-\frac{\pi^{2}(n-\langle n\rangle_{{\rm KZM}})^{2}}{6\langle n\rangle_{{\rm KZM}}}\right].
\end{eqnarray}

The experimental data shows deviations from the scaling regime at
short and long quench times that become more and more pronounced with
increasing order $q$ of the cumulant $\kappa_{q}$. Deviations at
long times are attributed to dephasing-induced heating and result
in an anti-KZM behavior whereby the value of $\kappa_{q}$ increases
with $\tau_{Q}$. Deviations at short times are of a different nature:
For $J\tau_{Q}\sim1$ deviations from the Landau-Zener formula occur,
as shown in Fig. 2 of the main text.

For the sake of clarity we elaborate on the connection between the
distribution of kink pairs and the distribution of the total number
of kinks. To account for the latter, we note that its characteristic
function is simply obtained as 
\begin{eqnarray}
\prod_{k>0}\left[1+\pap{e^{2i\theta}-1}p_{k}\right],
\end{eqnarray}
where the factor of $2\theta$ (instead of the $\theta$) accounts
for the creation of kinks in pairs. Cumulants $\kappa_{q}^{T}$ of
the total kink number distribution are simply related to cumulants
$\kappa_{q}$ of the distribution of pairs by the identity 
\begin{eqnarray}
\kappa_{q}^{T}=2^{q}\kappa_{q}.
\end{eqnarray}


\section{Experimental Setup}

The experimental task is to simulate the Hamiltonian of the $k$th mode
in Eq.~\eqref{H3K} using the trapped ion qubit. According the the time-dependent Schr\"odinger
equation, if we scale the time of evolution  as $t'=t\sin\left(ka\right)$,
the Hamiltonian will be change to 
\begin{align}
\hat{H}_{k}'& =\hat{H_{k}}/\sin\left(ka\right)=2J\frac{g-\cos\left(ka\right)}{\sin\left(ka\right)}\hat{\sigma}^{z}+2J\hat{\sigma}^{x}.\label{eq:H4K}
\end{align}
Eq.~\eqref{eq:H4K} can be taken as a Hamiltonian of single qubit driven by a microwave field
\begin{equation}
\hat{H}_{k}^{s}\pap{t}=\frac{\hbar}{2}\left(\Delta_{k}(t)\hat{\sigma}^{z}+\Omega_{R}\hat{\sigma}^{x}\right),\label{eq:H_qubit}
\end{equation}
where 
\begin{equation}
\Delta_{k}(t)=4J\frac{g\pap{t}-\cos\left(ka\right)}{\hbar\sin\left(ka\right)}\label{eq:detuning}
\end{equation}
is the detuning between the microwave frequency $\omega_{B}$ and the
qubit transition frequency $\omega_{Q}$, and where $\Omega_{R}=4J/\hbar$ is
the Rabi frequency under resonant microwave. In the experiment, the Rabi
frequency of the qubit simulator is set around 20 kHz, which depends
on the driven power of our microwave amplifier. Higher microwave power
can  shorten the Rabi time $T_{R}=1/\Omega_{R}$, and reduce
the total  operation time. To simulate the quench process, one would like to vary 
$g$ from $-\infty$ to 0, an idealized evolution that needs infinite time,
which can not be realized in experiment. However, to explore the universality associated with the crossing of the critical point  one can initialize the system sufficiently deep in the paramagnetic phase.
According to the KZM,  it is sufficient to  choose an initial value of the magnetic field such that the corresponding equilibrium relaxation time is much smaller than the time left until crossing the critical point. The system is then prepared out of the ``frozen region'' in the language of the adiabatic-impulse approximation \cite{Dziarmaga05}. 
For the quench time $\tau_{Q}\geq1$, the initial value of $g=-5$
is far out of the frozen time. We can simulate the TFQIM with an initial $g=-5$
and no  excitations, by preparing the  initial state of the qubit
 in the ground state of $\hat{H}_{k,i}^{s}=\hat{H}_{k}^{s}(g=-5)$
before the quench process. The initial state can be derived by solving
the eigenvectors of Eq.~\eqref{eq:H_qubit}, which is
\begin{equation}
\left|\psi_{-}\right\rangle _{k,i}=\cos\frac{\theta_{k,i}}{2}\left|0\right\rangle -\sin\frac{\theta_{k,i}}{2}\left|1\right\rangle ,\label{eq:phi_i}
\end{equation}
where $\theta_{k,i}=-\arctan\frac{\sin\left(ka\right)}{5+\cos\left(ka\right)}$
. 
\begin{figure}[t!]
	\centering \includegraphics[width=0.8\textwidth]{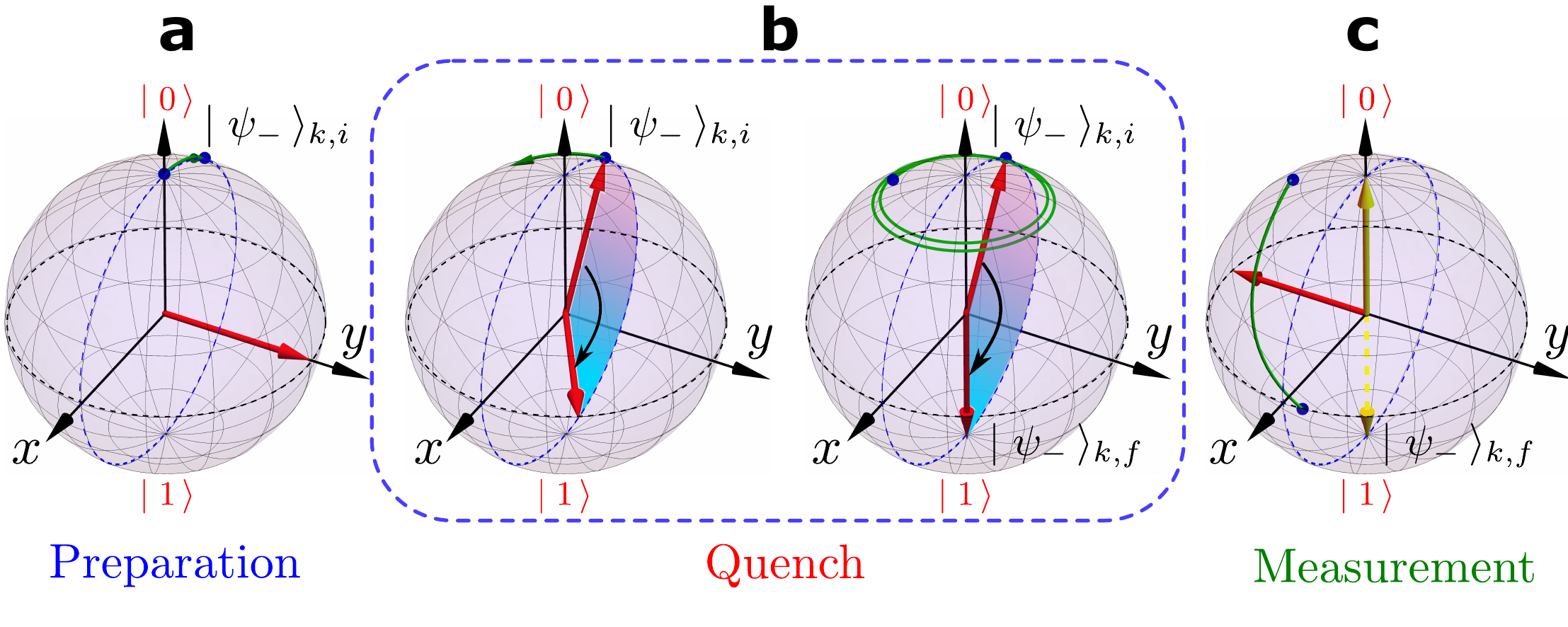} \caption{\textbf{Scheme to measure the excitation probability.} The
			quantum critical dynamics of the one-dimensional TFQIM is detected
			by measuring corresponding Landau-Zener crossings governing the dynamics
			in each mode. For each mode, a typical process to measure the excitation
			probability in three stages is shown in \textbf{a}, \textbf{b} and
			\textbf{c}.}
	\label{fig_block}
\end{figure}

The scheme to detected the quantum critical dynamics of the one-dimensional
TFQIM in each mode by using a single qubit is shown in Fig.\ \ref{fig_block}. 
The whole process can be divided to three steps. Before the process,
the qubit has been pumped to $\left|0\right\rangle $ state by using
a 369nm laser to excite transition ${S_{1/2}},\,F=1\rightarrow\ensuremath{{P_{1/2}},\,F=1}$.
In the first stage, the ion-trap qubit  is prepared into the state $\left|\psi_{-}\right\rangle _{k,i}$
by a resonant microwave pulse. The second stage is the quench process;
the Hamiltonian is time dependent and varies from $\hat{H}_{k,i}^{s}\equiv \hat{H}_{k}^{s}(g=-5)$
to $\hat{H}_{k,f}^{s}\equiv \hat{H}_{k}^{s}(g=0)$.  This is implemented   by
driving a chirped pulse into the qubit. The chirped pulse is in the
form of Eq. (\ref{eq:detuning}) with $g(t)=5(t/T_{p}-1)$, where
$T_{p}=5J\tau_{Q}\sin(ka)/\left(\hbar\Omega_{R}\right)$ is the pulse
length. The third stage is to measure the excitation probability after
the quench, which is to measure the occupation probability on the excited eigenstate
of $\hat{H}_{k,f}^{s}$. The excited eigenstate is $\left|\psi_{-}\right\rangle _{k,f}=\sin\theta_{k,f}\left|0\right\rangle +\cos\theta_{k,f}\left|1\right\rangle $,
where $\theta_{k,f}=-ka$. As the fluorescence detection on $^{171}\mathrm{Yb}^{+}$
can only discriminate $\left|0\right\rangle $ and $\left|1\right\rangle $
sates, we need to rotate the state $\left|\psi_{-}\right\rangle _{k,f}$
to $\left|1\right\rangle $ and then detect the bright state probability. 
We thus use a qubit rotation and a fluorescence detection  in this
stage. The calculated waveform to set the AWG to perform the pre-rotation,
quench and post-rotation in the three stages will be discussed in
the next section. 

\paragraph*{Driving Waveform with an Arbitrary Wave Generator:} We consider the driving microwave from AWG is $A\cos(\omega_{c}t+\varPhi(t))$,
where $A$ is the amplitude, $\omega_{c}$ is the carrier frequency
and $\varPhi(t)$ is an arbitrary phase function. The carrier frequency
can be mixed up with some frequency $\omega_{0}$ from a local oscillator
(LO) by using a frequency mixer. The waveform at the output of the
mixer is
\[
A\cos(\omega_{c}t+\varPhi(t))\cos(\omega_{o}t)=A/2\left(\cos((\omega_{o}+\omega_{c})t+\varPhi(t))+\cos((\omega_{o}-\omega_{c})t+\varPhi(t))\right).
\]
When the wave passes through the high pass filter, the wave form is filtered as $A/2\cos((\omega_{o}+\omega_{c})t+\varPhi(t))$. In the
experiment, we choose the frequency $\omega_{o}+\omega_{c}$ resonant
with the qubit transition $\omega_{Q}=\omega_{o}+\omega_{c}$, so the
magnetic field of the final driving waveform is $\mathbf{B}(t)=\mathbf{B_{0}}\cos(\omega_{Q}t+\varPhi(t))$.
The interaction between the magnetic field and spin is $H_{I}=-\mathbf{\mu}\cdot\mathbf{B}$,
where $\mu$ is the magnetic dipole of the spin qubit. The Hamiltonian
can be further simplified after the rotating wave approximation in the
interaction frame
\begin{equation}
\hat{H}_{I}(t)=\frac{\hbar}{2}\left[\Omega_{R}\left|1\right>\left\langle 0\right|e^{i\varPhi(t)}+\Omega_{R}^{*}\left|0\right>\left\langle 1\right|e^{-i\varPhi(t)})\right],\label{H4E}
\end{equation}
where the Rabi frequency $\Omega_{R}=-\left\langle 0\right|\mathbf{\mu}\cdot\mathbf{B}\left|1\right>/\hbar$.
The phase function $\varPhi(t)$ corresponds to the azimuthal angle
in the Bloch sphere. Equation \eqref{eq:H_qubit} can be also expressed
in interaction frame 
\[
H_{I,k}^{s}=\frac{\hbar}{2}\left(\Omega_{R}\left|1\right>\left\langle 0\right|\exp\left(i\int\Delta(t)dt\right)+\Omega_{R}^{*}\left|0\right>\left\langle 1\right|\exp\left(-i\int\Delta(t)dt\right)\right),
\]
so in the quench stage, the phase function is 
\[
\varPhi(t)=\intop_{0}^{T_{P}}\Delta(t)dt=\frac{\Omega_{R}}{\sin ka}\left(\frac{\Omega_{R}}{\tau_{Q}\sin ka}t^{2}-\left(5+\cos ka\right)t\right).
\]
We rotate the state along a vector in the equatorial plane to prepare
a state from $\left|0\right>$, or measure state to $\left|1\right>$,
which means $\varPhi(t)$ is constant in these two stages. The pulse
length for the preparation and measurement stages is determined by
the polar angle of the ground state at the beginning and end of the quench
process, $t=\theta/\Omega_{R}$, respectively. The whole expression
of $\varPhi(t)$ in the three stages can be derived as
\[
\varPhi(t)=\begin{cases}
\frac{\pi}{2}, & (0,\,t_{1})\\
\frac{\Omega_{R}}{\sin ka}\left(\frac{\Omega_{R}}{\tau_{Q}\sin ka}(t-t_{1})^{2}-\left(5+\cos ka\right)(t-t_{1})\right), & (t_{1},\,t_{2})\\
\phi_{f}-\frac{\pi}{2}, & (t_{2},t_{3})
\end{cases}
\]
where $t_{1}=(2\pi-\theta_{k,i})/\Omega_{R}$, $t_{2}=t_{1}+T_{p}$,
$t_{3}=t_{2}+(2\pi+\theta_{k,f})/\Omega_{R}$ and $\phi_{f}=-5\tau_{Q}(2.5+\cos ka)$.

\paragraph*{Qubit preparation and detection error:}

There are some limitations in the preparation and measurement of the
qubit. We measured the error through a preparation and detection experiment.
First, we prepare the qubit in the $\left|0\right>$ state by optical
pumping method, and detect the ion fluorescence. Ideally, no photon
can be detected as the ion is in the the dark state. However, there
are dark counts of the photon detector  as well as photons scattered
from the environment. We repeat the process $10^{5}$ times, and estimate
the detected photon numbers. We also repeat the process 1000 times
by preparing the qubit in the bright state. Histograms for the photon
number in the dark and bright states are shown in the Supplementary
Figure.~\ref{fig_SPM_error}. In the experiment, the threshold is
selected as 2: the state of the qubit is identified as the bright
state when the detected photon number is greater or equal to 2. Obviously,
there is a bright error $\epsilon_{B}$ for the dark state above the threshold,
and a dark error $\epsilon_{D}$ for the bright state below the threshold.
The total error can be take as $\epsilon=(\epsilon_{B}+\epsilon_{D})/2$.
\begin{figure}[h!]
	\centering \includegraphics[width=0.8\textwidth]{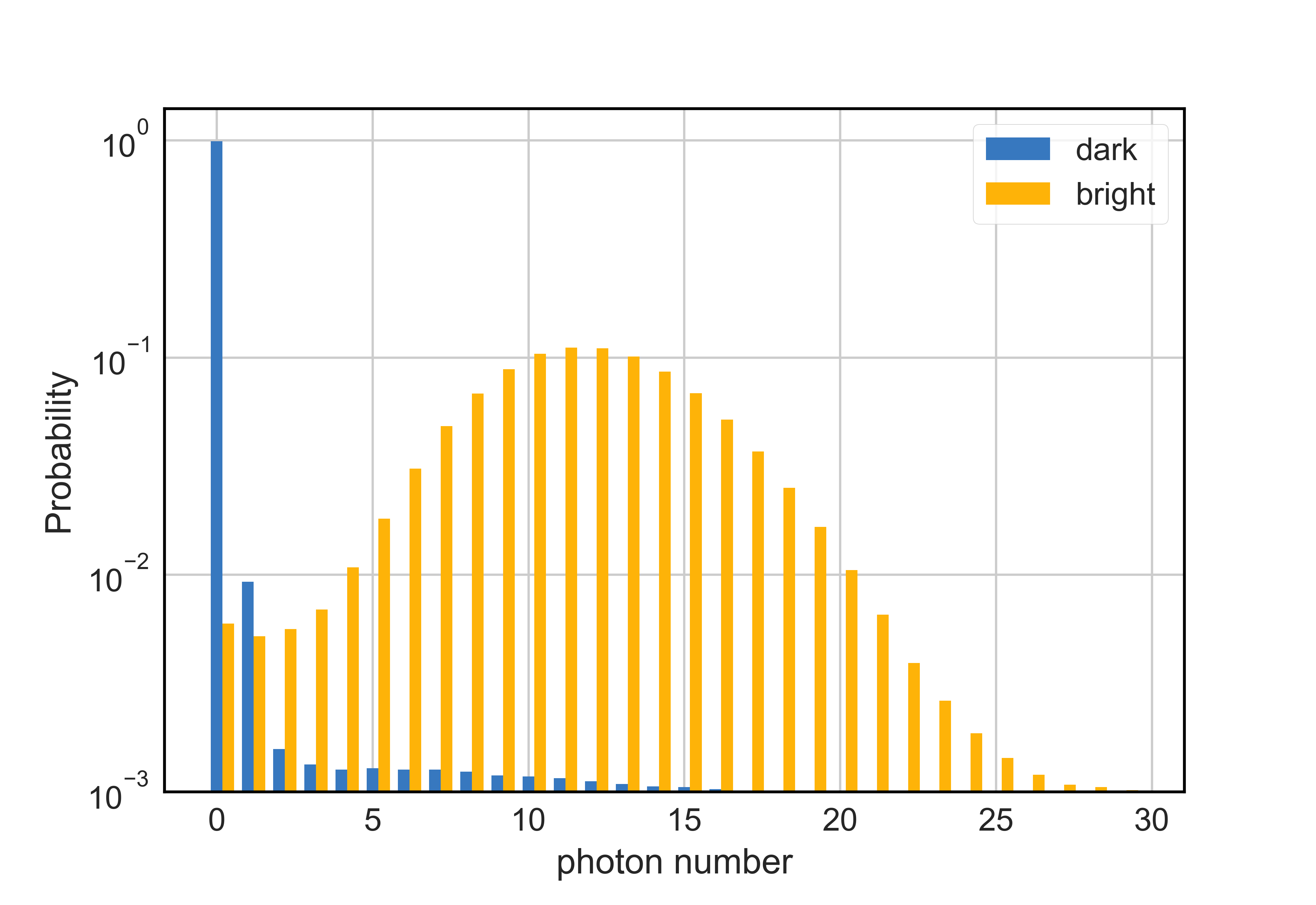} \caption{\textbf{Histogram for photon counts in state preparation and detection
			experiments.} The distribution of photon counts is shown when the
		qubit state is prepared in $\left|0\right>$ (dark state) and $\left|1\right>$
		(bright state). }
	\label{fig_SPM_error} 
\end{figure}

\end{document}